# Superstatistics and temperature fluctuations

*F. Sattin[1]*

*Padova, Italy*

**Abstract**

Superstatistics [C. Beck and E.G.D. Cohen, Physica A **322**, 267 (2003)] is a formalism aimed at describing statistical properties of a generic extensive quantity *E* in complex out-of-equilibrium systems in terms of a superposition of equilibrium canonical distributions weighted by a function *P(β)* of the intensive thermodynamic quantity β conjugate to *E*. It is commonly assumed that *P(β)* is determined by the spatiotemporal dynamics of the system under consideration. In this work we show by examples that, in some cases fulfilling all the conditions for the superstatistics formalism to be applicable, *P(β)* is actually affected also by the way the measurement of *E* is performed, and thus is not an intrinsic property of the system.

**1. Introduction**

Superstatistics is a term coined and introduced in statistical thermodynamics a few years ago by Beck and Cohen [1,2] with the purpose of modelling non Maxwell-Boltzmann statistical distributions in complex systems out of equilibrium. It provides potentially a rationale for reconciling an ample variety of empirical distributions under one single concept; this explains the popularity enjoyed: superstatistics has been employed in nuclear physics, turbulent fluids, high-energy physics, plasmas and gravitational systems, to mention a few (a search over Internet bibliographic databases with the "superstatistics" keyword returned nearly 300 papers by the end of 2017, while the original paper [1] had more than 700 Google citations).

Superstatistics claims that in systems where fluctuations of temperature do exist, coarse-grained measurements of energy, performed over spatial and temporal scales larger than those defined by the correlation properties of the temperature, yield statistical distributions that can be written as superposition of canonical distributions:

$$P_S(E) = \int d\beta \, P_\beta(\beta) P_{MB}(E) \, \rho(E)/Z(E) \qquad P_{ME}(E) = \exp(-\beta E) \qquad (1)$$

In (1) *E* is the energy and β the inverse temperature [In the following we will adhere to the convention of naming the extensive variable *E* the "energy", and the intensive 1/β the "temperature", although they may have broader meanings, of course]; ρ(*E*) the density of states with energy *E,* and *Z* the normalization term. The function $P_\beta(\beta)$ is understood to be a weight function, whose meaning is to quantify the probability for the system to be in a state of local thermal equilibrium with temperature 1/β. As such, it is expected to be an intrinsic property of the system studied, independent of any detail of the measurement of the conjugate variable *E* as long as it is carried out over large enough scales. The validity of this hypothesis is of relevance in experimental applications, where one has information about $P_S(E)$ and uses (1) to infer $P_\beta(\beta)$ and gain knowledge about the system' statistical properties [3,4].

---

[1] Email: fabio.sattin@fastwebnet.it; Permanent address: Consorzio RFX, Corso Stati Uniti 4, Padova, I-35127

In this brief note I will show that, rather surprisingly, this hypothesis is not always true. Specifically, I design two computer models, variants of each other, where temperature features fluctuations with a given statistical pattern quantified by a distribution $f(\beta)$, and build a synthetic measurement of the coarse-grained distribution $P_S(E)$, which is then projected onto the canonical basis function according to Eq. (1). Analytical and numerical arguments will show that, in one case, the actual statistical distribution $f(\beta)$ and the inferred one $P_\beta(\beta)$ do match: $P_\beta(\beta) = f(\beta)$, as naively expected. The second version of the model, instead, yields $P_\beta(\beta) \neq f(\beta)$. It is explicitly shown that the function $P_\beta(\beta)$ does not depend from the properties of the medium alone, but rather is crucially determined by the way the measurement of $P_S(E)$ is performed.

## 2. Description of the models

Let us consider the well-known case of a 1-dimensional Brownian particle and write accordingly its Langevin equation:

$$\dot{v} = -\gamma v + wL(t) \qquad <L(t)> = 0, \quad <L(t)L(t')> = \delta(t-t'), \quad w^2 = \frac{2\gamma}{\beta} \qquad (2)$$

Hereafter, γ will be considered as constant. Instead, $w$ (or, equivalently, β) is taken as randomly varying in space and/or time. Specifically, we consider the following two variants.

*Model 1*

The inverse temperature β is picked up from a uniform distribution within the interval $(\beta_{\min}, \beta_{\max})$ and is held constant during a time interval Δ chosen either constant: $\Delta = \Delta_c$, or randomly picked up from the uniform distribution $(\Delta_{\min}, \Delta_{\max})$. Both $\Delta_c$ and $\Delta_{\min}$ are chosen large enough so that the particle has the time to thermalize. At the end of each time interval a new β value is picked up and the numerical integration of Eq. (2) continued. We track one particle over very long times and record both its velocity and the instantaneous value of β at equispaced steps. The numerical integration of the Langevin equation (2) is performed using the algorithm described in [5].

*Model 2*

The space is divided into cells of length λ. Within each cell the inverse temperature β is picked up independently and randomly from a uniform distribution within the interval $(\beta_{\min}, \beta_{\max})$. The length λ is large enough to allow for thermalization of the particle inside the cell. Again, we sample the particle velocity at equispaced intervals in order to build energy and beta distribution.

## 3. Results

The statistical distribution of inverse temperatures is

$$f(\beta) = \frac{1}{\beta_{max}-\beta_{min}} \times \begin{cases} 1 & \beta_{min} < \beta < \beta_{max} \\ 0 & otherwise \end{cases} \qquad (3)$$

We insert (3) into (1). For convenience, hereafter we will work using velocities rather than energies; thus (1) becomes

$$P_S(u) = \frac{1}{\beta_{max}-\beta_{min}} \int_{\beta_{min}}^{\beta_{max}} d\beta \exp\left(-\frac{\beta u^2}{2}\right) \sqrt{\frac{\beta}{2\pi}} = \frac{1}{\beta_{max}-\beta_{min}} \frac{1}{u^3} \left\{ \mathrm{erf}\left(\sqrt{\frac{\beta_{max}}{2}} u\right) - \mathrm{erf}\left(\sqrt{\frac{\beta_{min}}{2}} u\right) - u\sqrt{\frac{2}{\pi}} \left[\sqrt{\frac{\beta_{max}}{2}} e^{-\frac{\beta_{max} u^2}{2}} - \sqrt{\frac{\beta_{min}}{2}} e^{-\frac{\beta_{min} u^2}{2}} \right] \right\} \qquad (4)$$

In Fig. (1) we show the PDF($u$) as sampled from the Brownian particle motion according to model 1 alongside the theoretical expectation (4). Parameters are: $\beta_{min} = 0.25, \beta_{max} = 4.0, \gamma = 0.01$, $\Delta_{min} = 3 \cdot 10^5$, $\Delta_{max} = 8 \cdot 10^5$. Time step for integration of the Langevin equation is $dt = 1$. The state of the particle is saved every $10^5$ time steps.

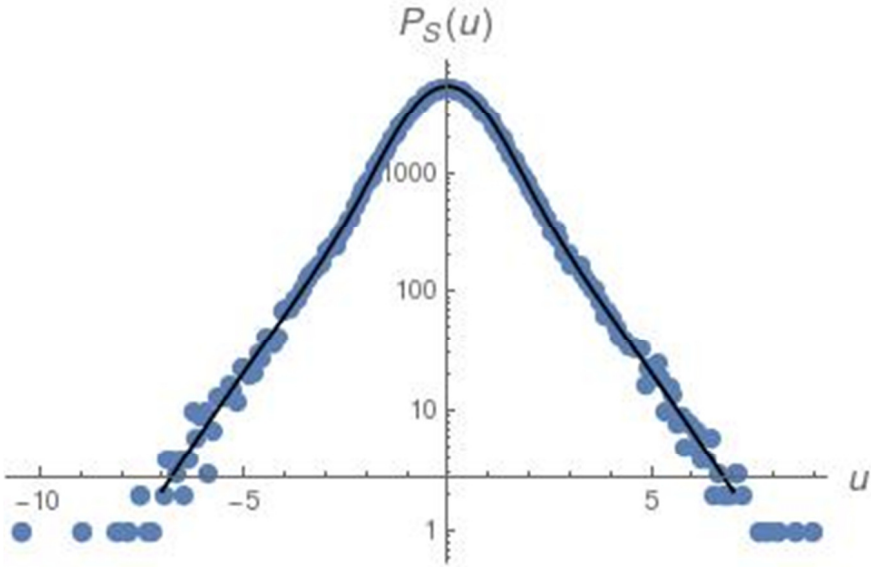

**Fig. 1**. Dots, sampled $u$ distribution for model 1. Solid curve, Eq. 4.

There is nice agreement. Furthermore, in Fig. (2) we show the distribution of β the way it has actually been sampled by the particle. Consistently with our expectations, we recover back the uniform distribution (3).

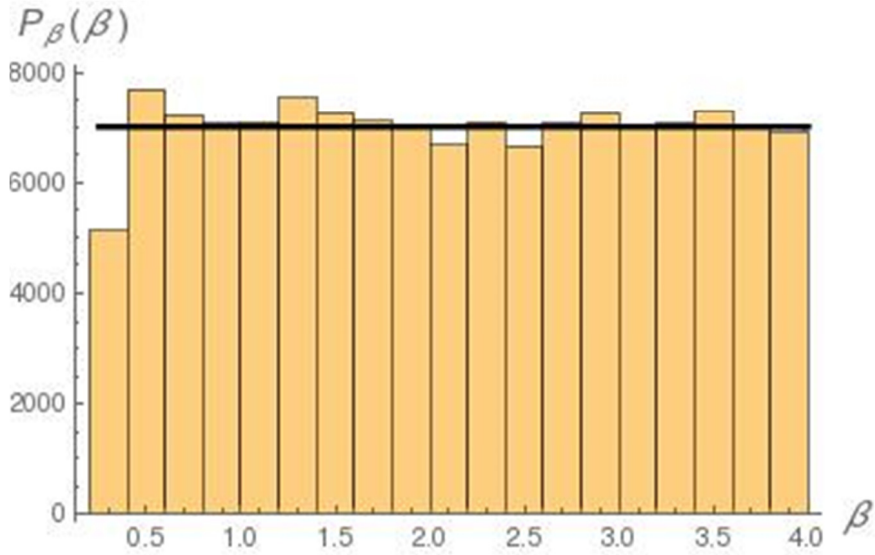

**Fig. 2**. Histogram of sampled β values for model 1. The horizontal straight line is a guide to the eye.

Let us move to model 2. The parameters used are now: $\beta_{\min} = 0.25, \beta_{\max} = 4.0, \gamma = 0.01$. Time step for integration of the Langevin equation is $dt = 1$. The box width is set to $\lambda = 10^3$. The state of the particle is saved every $10^5$ time steps.

This time we show first the sampled distribution for β (Fig. 3). It is definitely different from the shape (3): it is no longer uniform, rather features a linear trend.

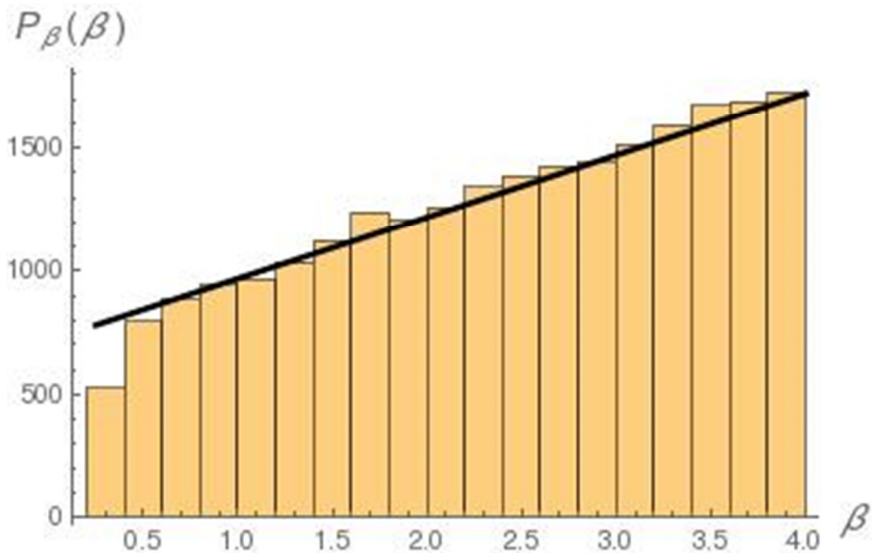

**Fig. 3**. Histogram of sampled β values for model 2. The black curve is a straight line used to guide the eye.

This is easily explainable: the dynamics a Brownian particle thermalized in the $i$-th cell with local inverse temperature $\beta_i$ is described by a Fokker-Planck equation with diffusion coefficient $D \propto \beta_i^{-1}$, and the average time spent by the particle in order to exit the cell is $\tau \sim \frac{\lambda^2}{D} \sim \beta_i$. Ultimately, therefore, the particle is embedded within each thermal bath with inverse temperature β for a fraction of time that scales as β. If, rather than (3), we insert into Eq. (1) the effective distribution

$$f'(\beta) = \frac{2}{\beta_{max}^2 - \beta_{min}^2} \times \begin{cases} \beta & \beta_{min} < \beta < \beta_{max} \\ 0 & otherwise \end{cases} \tag{5}$$

we get

$$P_S'(u) = \frac{2}{\beta_{max}^2 - \beta_{min}^2} \int_{\beta_{min}}^{\beta_{max}} d\beta \, \beta \exp\left(-\frac{\beta u^2}{2}\right) \sqrt{\frac{\beta}{2\pi}} =$$

$$\frac{1}{\beta_{max}^2 - \beta_{min}^2} \sqrt{\frac{2}{\pi}} \frac{1}{u^5} \left\{ 3\sqrt{2\pi} \left[ \mathrm{erf}\left(\sqrt{\frac{\beta_{max}}{2}}\, u\right) - \mathrm{erf}\left(\sqrt{\frac{\beta_{min}}{2}}\, u\right) \right] - 2u \left[ \sqrt{\beta_{max}}\, e^{-\frac{\beta_{max} u^2}{2}} (3 + \beta_{max}\, u^2) - \sqrt{\beta_{min}}\, e^{-\frac{\beta_{min} u^2}{2}} (3 + \beta_{min}\, u^2) \right] \right\} \tag{6}$$

In Fig. (4) we plot the empirical distribution alongside both guesses (4) and (6). The superiority of the latter in interpolating the data is obvious.

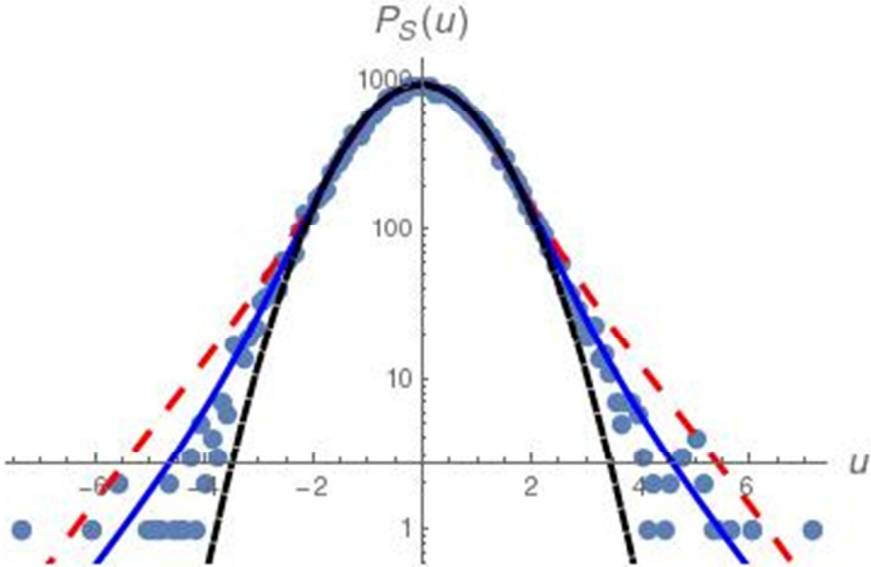

**Fig. 4.** Histogram of the sampled velocity *u* (dots) for model 2. The curves are: a gaussian (short-dashed black curve); Eq. 4 (red long dashed curve); Eq. 6 (solid blue curve).

## 4. Conclusions

The result of this simple exercise can be summarized in a caveat: the core concept of superstatistics, that the behavior of several non-equilibrium systems, sampled over suitably large time and space scale, can be compactly modeled as the superposition of statistics coming from systems at different local equilibria, is useful, but care must be exerted when attempting to infer system properties from the weighting function $P_\beta(\beta)$, since there is no guarantee at all *a priori* that it be not affected also by the way the measurement of the conjugate quantity *E* is performed. Since an experimentalist is not expected to have access to the details of the fluctuating system, when collecting $P_\beta(\beta)$ from experiments, she cannot tell whether it arises from a model of the kind 1 or 2; thus, one and the same inferred distribution may be produced by actually distinct microscopic statistics.

**Acknowledgments**

M. Baiesi is acknowledged for reading the manuscript.